\documentclass[epsfig,10pt,twocolumn]{article}
\usepackage{epsfig,citesort,setspace}
\usepackage{times}
\input{amssym}

\begin{document}
\title{\bf \Large Order independent structural alignment of circularly permuted proteins}
\author{\vspace*{-.5in}
	 $^1$T. Andrew Binkowski, 
               $^2$Bhaskar DasGupta\thanks{Supported by NSF grants: CCR-0296041, CCR-0206795 and CCR-0208749. }, 
          and 
               $^1$Jie Liang\thanks{Supported by NSF grants: CAREER DBI-0133856,  DBI-0078270, and  NIH grant: GM-68958.}}
\date{\vspace*{-.1in} Departments of $^1$Bioengineering and $^2$Computer Science,The University of Illinois at Chicago, IL, 60305\\
\vspace*{.2in}
{\Large Accepted by {\it 
{\it Proc.\ 26th IEEE EMBC Conference, San Francisco, 2004}. 
}}
}
\maketitle

\setcounter{page}{1}

{\small {\em Abstract}--{\bf Circular permutation connects the N and C
termini of a protein and  concurrently cleaves elsewhere in the chain,
providing an important mechanism for generating novel protein fold and
functions.   However,  their in  genomes  is  unknown because  current
detection methods  can miss many occurances,  mistaking random repeats
as  circular permutation.   Here  we develop  a  method for  detecting
circularly  permuted proteins  from  structural comparison.   Sequence
order independent alignment of protein structures can be regarded as a
special case  of the maximum-weight independent set  problem, which is
known   to  be   computationally  hard.    We  develop   an  efficient
approximation  algorithm  by  repeatedly  solving  relaxations  of  an
appropriate intermediate integer programming formulation, we show that
the approximation ratio is much better then the theoretical worst case
ratio  of  $r  =  1/4$.   Circularly  permuted  proteins  reported  in
literature  can be  identified  rapidly with  our  method, while  they
escape  the detection  by  publicly available  servers for  structural
alignment.}}
\vspace{0.6em}

{\small    {\em   Keywords}--{\bf   circular    permuations,   integer
programming, linear programming, protein structure alignment}}
\vspace{0.6em}
\vspace*{-.25in}
\subsubsection*{\center {\sc Introduction}}
A  circularly permuted protein  arises from  ligation of  the N  and C
termini of  a protein and  concurrent cleavage elsewhere in  the chain
\cite{Lindqvist_1997,RussellPonting_98}.  
In  nature, circular  permutation  often originate  from
tandem  repeats  via  duplication  of  the C-terminal  of  one  repeat
together with  the N-terminal of  the next repeat,  as is the  case of
swaposin.   Another mechanism is ligation
and cleavage of peptide chains during post-translational modification,
as is the case of concanavalin A.
The  full   extent  of  circular  permutation   and  thier  biological
consequences are currently unknown.  Discovery of circular permutation
at  genome   wide  scale  will   enable  systematic  studies   of  its
contribution  to the generation  of novel  protein function  and novel
protein fold.

Currently,  sequence alignment  based methods  are the  only available
tools   for   detecting    circular   permutations.    These   include
postprocessing  output from standard  dynamic programming  methods, as
well  as  customized  algorithms  \cite{Uliel_1999}.   Sequence  based
methods can miss many circularly permuted proteins, because either one
or both fragments  may escape detection by local  alignment if the two
proteins are distantly related.

Protein structures are far more conserved than sequences, and they can
reveal very distant evolutionary relationship \cite{pvsoar03}.  Jung
et.   al. \cite{Jung01_PS} showed  that there  is potential  to discover
remotely related  protein domains by  artifically permutating proteins
and  superimposing them  on native  domains in  the  protein structure
database.   Detecting  circular permutation  from  structures has  the
promise to  uncover many more  ancient permutation events  that escape
sequence methods.

In  this study,  we describe  a new  algorithm for  detecting circular
permutation  in protein  structures. We  show with  examples  that our
algorithm can align circular  permutations reported in literature.  
Our work introduces an efficient approximate method for protein
substructure comparison. Two protein structures are first cut into
pieces exhaustively of varying lengths and 
then compared. An approximation
algorithm is used to search for optimal combination of peptide pieces
from both structures.
With the special nature of spatial distribution of proteins,
a fractional version of local-ratio approach for scheduling
split-interval graphs works well for detecting very similar spatial
substructures. 
Our experimentation showed that the
approximation ratio is excellent, with an average value of at least
$1/2.53$.

\subsubsection*{\center {\sc Structural Alignment of Circularly Permuted Proteins}}
\begin{figure}[tb]
       \centerline{\epsfig{figure=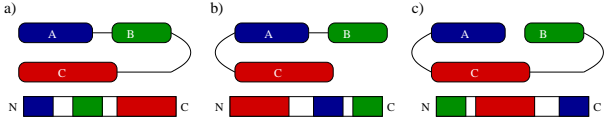,width=80mm}} 
\caption{\sf \footnotesize Three cartoon protein structures
       related by a circular permutation.  The location of domains
       A,B,C are shown in a primary sequence layout below each
       structure.}
       \label{Fig:3_permutations}
\end{figure}

Three  simplified  protein structures  that  are  related by  circular
permutation are  shown in Figure~\ref{Fig:3_permutations}.   The three
corresponding domains (labeled A, B, and C, respectively) are all very
similar  across proteins.   The global  spatial arrangements  of these
three domains are also very  similar.  However, the orderings of these
domains in primary sequence are completely different.

The discontinuity and different ordering of spatially neighboring
domains in primary sequence makes the detection of global structural
similarity.  Most structural alignment methods are unable to connect
the break where circular permutation occurs, and would prematurely
terminates the alignment.  In addition, the reverse ordering of
residues within a domain between two proteins also poses challenge for
such methods.  For example, the A domain in
Figure~\ref{Fig:3_permutations}a is ordered from N to C.  The same
domain in Figure~\ref{Fig:3_permutations}b is ordered from C to N.
The algorithm outlined below is well-suited to solve these types of
challenging problems.

\paragraph{\it Basic Methodology}
\newtheorem{defn}{Definition}
\newtheorem{lemma}{Lemma}[section]
\newcommand{\cS}{\mathcal{S}}
\newcommand{\vecyint}{\stackrel{\textstyle\longrightarrow}{y_{\rm int}}}
\newcommand{\vecyfrac}{\stackrel{\textstyle\longrightarrow}{y_{\rm frac}}}
\newcommand{\wonevec}{\stackrel{\textstyle\rightarrow}{w_1}}
\newcommand{\wtwovec}{\stackrel{\textstyle\rightarrow}{w_2}}
\newcommand{\wvec}{\stackrel{\textstyle\rightarrow}{w}}
\newcommand{\xvec}{\stackrel{\textstyle\rightarrow}{x}}
\newcommand{\dvec}{\stackrel{\textstyle\rightarrow}{d}}
We first introduce some notations: A substructure $\lambda_a$ of a
protein structure $\cS_a$ is a continuous fragment $\lambda_a= (a_1,
a_2) \in \Lambda_a$, where $a_1,a_2$ are the residue
numbers of the beginning and end of the substructure, respectively, and
$1 \le a_1 < a_2\le |\cS_a|$.  A residue $t \in \lambda_a$
is part of substructure $\lambda_a$ if $a_1 \le t \le a_2$.
$\Lambda_a$ is the set of all possible continuous substructures or
fragments of protein structure $\cS_a$, {\it i.e.}, $\Lambda_a =
\{(a_1, a_2) \}$. $\cS_b, \lambda_b, b_1, b_2$, and $\Lambda_b$ are
similarly defined.

The  problem of  protein substructure  comparison is  captured  in the
following   {\em   Basic   Substructure   Similarity   Identification}
(BSSI$_{\Lambda,\sigma}$) problem.

{\footnotesize 
\begin{defn}
{\rm
\begin{description}
\item[Instance:] 
a set $\Lambda\subseteq \Lambda_a\times \Lambda_b$ of ordered pairs
of substructures of $\cS_1$ and $\cS_2$
and a {\em similarity} function
$\sigma:\Lambda\rightarrow{\Bbb
R_+}$ mapping these pairs of substructures
to similarity values (non-negative real numbers).

\item[Valid solutions:] a set
$\{
\lambda_1, \lambda_2, \cdots, \lambda_k
\}
=
\{(\lambda_{a,1},\lambda_{b,1}),
(\lambda_{a,2},\lambda_{b,2}),\cdots, (\lambda_{a,k},\lambda_{b,k})\}$
of pairs of substructures such that $\lambda_{\ell,i} \in
\Lambda_\ell$ and $\lambda_{\ell,j} \in \Lambda_\ell$ are disjoint for
$\ell\in\{a,b\}$ and $i\neq j$.

\item[Objective:]
{\em maximize} the
total similarity of the selection
$\sum_{i=1}^k\sigma(\lambda_{a,i},\lambda_{b,i})$.
\end{description}
}
\end{defn}
}

The BSSI$_{\Lambda,\sigma}$
problem is a special case of the famous
maximum-weight independent set (MWIS) problem in graph theory \cite{AFWZ95}.
BSSI$_{\Lambda,\sigma}$
is MAX-SNP-hard even when the substructures are
restricted to  short lengths~\cite{YHNSS02}\footnote{A maximization problem being
MAX-SNP-hard implies that there exists a constant $0<\varepsilon<1$ such that
no polynomial time algorithm can return a solution with a value of
at least $(1-\varepsilon)$ times the optimum value.}. 
Our approach 
is to adopt the approximation
algorithm for scheduling split-interval graphs~\cite{YHNSS02},
which is based on a fractional version of the local-ratio
approach.
We introduce the following 
definitions:
{\small
\begin{itemize}
\item
The conflict graph $G_{\Delta}=(V_{\Delta},E_{\Delta})$
for any subset $\Delta\subseteq\Lambda$,
is a graph in which $V_{\Delta}=\Delta$ and
$E_{\Delta}$ consists of all distinct pairs of vertices
$\{(\lambda_a,\lambda_b),(\lambda_a',\lambda_b')\}$
from $V_{\Delta}$ such that $\lambda_i$ and $\lambda_i'$
are {\em not} disjoint for some $i\in\{a,b\}$.

\item
The closed neighborhood of a vertex $v$ of $G_\Delta$,
denoted by Nbr$_\Delta[v]$,
is defined as $\{u\;|\;\{u,v\}\in E_\Delta\}\cup\{v\}$.
\end{itemize}
}
The recursive algorithm for solving
BSSI$_{\Lambda,\sigma}$
is as follows:
\begin{table}[tb]
\begin{footnotesize}
\caption{\label{Tab:lpformulation} \sf \footnotesize The LP formulation of the BSSI problem.}
\vspace*{0.1in}
\begin{tabular}{|lll|}\hline
& & \\
\multicolumn{3}{|l|}{{\em Maximize}
$\;\;\;\;\sum_{\lambda\in\Lambda}\sigma(\lambda)\cdot x_\lambda$} \\
& & \\
{\em subject to} & & \\
& & \\
\hspace*{0.1in}
$\displaystyle\sum_{t\in\lambda_a\in\Lambda_a\;\&\;\lambda=(\lambda_a,\lambda_b)\in\Lambda} 

y_{\lambda,\lambda_a}\leq 1$ &
$\forall t\in \Lambda_a$ & {\bf (1)} \\
& & \\
\hspace*{0.1in}
$\displaystyle\sum_{t\in\lambda_b\in\Lambda_b\;\&\;\lambda=(\lambda_a,\lambda_b)\in\Lambda} 

y_{\lambda,\lambda_b}\leq 1$ &
$\forall t\in \Lambda_b$ & {\bf (2)} \\
& & \\
\hspace*{0.1in}
$y_{\lambda,\lambda_a}-x_\lambda\geq 0$ &
$\forall\lambda=(\lambda_a,\lambda_b)\in\Lambda$ & \multicolumn{1}{r|}{\bf 
(3)} \\
\hspace*{0.1in}
$y_{\lambda,\lambda_b}-x_\lambda\geq 0$ &
$\forall\lambda=(\lambda_a,\lambda_b)\in\Lambda$ & \multicolumn{1}{r|}{\bf 
(4)} \\
\hspace*{0.1in}
$x_\lambda,y_{\lambda,\lambda_a},y_{\lambda,\lambda_b}\geq 0$ &
$\forall\lambda=(\lambda_a,\lambda_b)\in\Lambda$ & \multicolumn{1}{r|}{\bf 
(5)} \\
& & \\ \hline
\multicolumn{3}{c}{} \\
\end{tabular}
\end{footnotesize}
\end{table}

{\small
\begin{itemize}
\item
Remove every substructure pairs $\lambda=(\lambda_a,\lambda_b)\in\Lambda$ such that
$\sigma(\lambda_a,\lambda_b)\leq 0$. If $\Lambda=\emptyset$
after these removals, then return $\emptyset$ as
the solution.

\item
Solve a linear programming (LP) formulation of the
BSSI$_{\Lambda,\sigma}$ problem by relaxing a corresponding integer
programming version of the BSSI$_{\Lambda,\sigma}$ problem. For every
$\lambda=(\lambda_a,\lambda_b)\in\Lambda$, introduce three indicator
variables $x_\lambda$, $y_{\lambda,\lambda_a}$ and
$y_{\lambda,\lambda_b} \in \{0,1\}$, but relaxed to real numbers. The
LP formulation is shown in Table~\ref{Tab:lpformulation}.
\end{itemize}

\begin{itemize}
\item
For every vertex $\lambda\in\Lambda$ of $G_\Lambda$, compute
its {\em local conflict number}
$\displaystyle\alpha_{\lambda,\Lambda,\sigma}=
\sum_{\mu\in\mbox{Nbr}_\Lambda[\lambda]}x_{\mu}$.
Let $\lambda_{\rm min}$ be a vertex with the minimum local
conflict number
$r_{\Lambda,\sigma}=\min_{\lambda}\{\alpha_{\lambda,\Lambda,\sigma}\}$.
Define a new similarity function
$\sigma_{\rm new}$ from $\sigma$ as follows
$
\sigma_{\rm new}(\lambda)=\left\{
\begin{array}{ll}
\sigma(\lambda), & \mbox{if $v\not\in\mbox{Nbr}_\Lambda[\lambda_{\rm min}]$} \\
\sigma(\lambda)-\sigma(\lambda_{\rm min}), & \mbox{otherwise} \\
\end{array}
\right.
$.

\item
Recursively solve the
BSSI$_{\Lambda,\sigma_{\rm new}}$ problem using the same approach.
Let $\Lambda'\subseteq\Lambda$ be the solution returned.

\item
If the substructure pair $\lambda_{\rm min}$ can be selected together with all
the pairs in $\Lambda'$ then return $\Lambda'\cup\{\lambda_{\rm min}\}$ as the
solution, else return $\Lambda'$ as the solution.
\end{itemize}
}
A brief explanation of the various inequalities in the LP formulation as
described above is as follows:
{\small
\begin{itemize}
\item
The {\it interval clique inequalities\/} in {\bf (1)} (resp.\ {\bf (2)}) ensures that
the various substructures of $\cS_a$ (resp. $\cS_b$) in the selected pairs of
substructures from $\Lambda$ are mutually disjoint.

\item
Inequalities in {\bf (3)} and {\bf (4)} ensure consistencies between
the indicator variable for substructure pair $\lambda$ and its two
substructures $\lambda_a$ and $\lambda_b$.

\item
Inequalities in {\bf (5)} ensure non-negativity of the indicator variables.
\end{itemize}
}
In implementation, the graph $G_\Delta$ 
is considered implicitly via intersecting intervals. The interval
clique inequalities can be generated via a {\em sweepline}
approach. The running time depends on the number of
recursive calls and the times needed to solve the LP formulations. Let
LP$(n,m)$ denote the time taken to solve a linear programming problem
on $n$ variables and $m$ inequalities. Then the worst-case running
time of the above algorithm is
$O(|\Lambda|\cdot\mbox{LP}(3|\Lambda|,5|\Lambda|+|\Lambda_1|+|\Lambda_2|))$.
However, the worst-case time complexity happens under the excessive
pessimistic assumption that each recursive call removes exactly one
vertex of $G_{\Delta}$, namely $\lambda_{\rm min}$ only, from
consideration, which is unlikely to occur in practice as our
computational results show.

The performance ratio of our algorithm is as follows.
Let $r$ be the maximum
of all the $r_{\Lambda,\sigma}$'s in all the recursive calls.
Proofs in~\cite{YHNSS02} translate to the fact that $r\leq 4$ and
the above algorithm returns a solution whose total similarity
is at least $1\over r$ times that of the optimum.
The value of $r$
is much smaller compared to $4$ in practice ({\it e.g.}, $r=2.53$).

\begin{figure}[tb]
       \centerline{\epsfig{figure=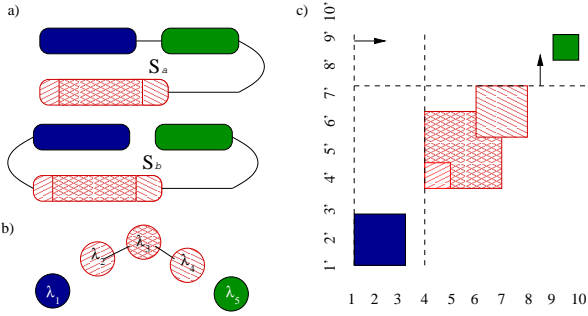,width=80mm}}    \caption{\sf
       \footnotesize   A  simplified   representation  of   the   {\em  Basic
       Substructure Similarity Identification} problem: a) The cartoon
       representation  of  circularly  permuted proteins  $\cS_a$  and
       $\cS_b$, b) The problem represented as a graph where each node,
       $\lambda$, represents  an aligned  fragment pair and  each edge
       represents a conflict, where the same residues are contained in
       different  fragments and  c)  An illustration  how sweep  lines
       (dashed)   can   identify    overlapping   aligned   pairs.   }
       \label{Fig:conflict_1}
\end{figure}

\paragraph{\it Implementation and Computational Details.}
We  present   a  simplified  example  for   illustration  two  protein
structures   $\cS_a$    (Figure~\ref{Fig:conflict_1}a)   and   $\cS_b$
(Figure~\ref{Fig:conflict_1}a)  are   selected  for  alignment.   Here
$\cS_b$  is the  structure to  be aligned  to the  reference structure
$\cS_a$.  We systematically cut $\cS_b$ into fragments of length 7--25
and  exhaustively compute  a similarity  score of  each  fragment from
$\cS_b$ to  all possible fragments  of equal length in  $\cS_a$.  Each
fragment pair can  be thought of as  a vertex in a graph,  as shown in
Figure~\ref{Fig:conflict_1}b.

Suppose we have the following similarity scores for aligned substructures:
{\footnotesize
$ \sigma(\lambda_1) = \sigma( (1,3)(1',3') )= 8,  
 \sigma(\lambda_2) = \sigma( (4,5)(4',5') )= 5, 
 \sigma(\lambda_3) = \sigma( (4,7)(4',7') )= 7,  
 \sigma(\lambda_4) = \sigma( (6,8)(6',8') )= 3, \mbox{ and } 
 \sigma(\lambda_5) = \sigma( (9,10)(9',10') )= 6.$  }

We can describe the problem  of selecting the best structural fragment
pairs as to maximize $8x_{\lambda_1}+5x_{\lambda_2}+7x_{\lambda_3}+3x_{\lambda_4}+6x_{\lambda_5} $.

Figure~\ref{Fig:conflict_1}b shows  the conflict graph for  the set of
fragments.     A   sweep    line   (shown    as   dashed    lines   in
Figure~\ref{Fig:conflict_1}c)  is implicitly constructed  ($O(n)$ time
after sorting)  to determine which vertices of  fragment pair overlap.
A conflict  is shown in Figure~\ref{Fig:conflict_1}b  as edges between
vertices.  Vertices  $\lambda_1$ and $\lambda_5$ do  not conflict with
any other  fragments, while $\lambda_2$ and  $\lambda_4$ conflict with
$\lambda_3$.   For   this  graph,   the  constraints  in   the  linear
programming formulation are shown in Table~\ref{Tab:constraints}.  The
linear  programming problem is  solved using  the {\sc  Bpmpd} package
\cite{Meszaros96_CMA}.

\begin{table}[bt]
\begin{footnotesize}
\begin{center}
\caption{\label{Tab:constraints} \sf \footnotesize The constraints of the consflict graph for the set of fragments in Figure~\ref{Fig:conflict_1}c.}
\vspace*{0.1in}
\begin{tabular}{|lll|}\hline
 Interval Clique inequality: & &{\bf (1)}\\ 
 $y_{\lambda_1,\lambda_a}\leq1$             & Line sweep at 1 &\\ 
 $y_{\lambda_2,\lambda_a}+y_{\lambda_3,\lambda_a}\leq1$ & Line sweep at 4, 5 &\\
 $y_{\lambda_3,\lambda_a}+y_{\lambda_4,\lambda_a}\leq1$ & Line sweep at 6, 7 &\\
 $y_{\lambda_4,\lambda_a}\leq1$              & Line sweep at 8  &\\
 $y_{\lambda_5,\lambda_a}\leq1$             & Line sweep at 9   &\\
& & \\
 Interval Clique inequality: & &{\bf (2)} \\
 $y_{\lambda_1,\lambda_b}\leq1$             & Line sweep at 1' & \\
 $y_{\lambda_2,\lambda_b}+y_{\lambda_3,\lambda_b}\leq1$ & Line sweep at 4', 5' &\\
 $y_{\lambda_3,\lambda_b}+y_{\lambda_4,\lambda_b}\leq1$ & Line sweep at 6', 7' &\\
 $y_{\lambda_4,\lambda_b}\leq1$             & Line sweep at 8' &\\
 $y_{\lambda_5,\lambda_b}\leq1$             & Line sweep at 9' &\\
& & \\

 Interval Clique inequalities: & & {\bf (3, 4)}\\
 $y_{\lambda_1,\lambda_a} - y_{\lambda_1}\geq0$, &  $y_{\lambda_1,\lambda_b} - y_{\lambda_1}\geq0$ &\\
 $y_{\lambda_2,\lambda_a} - y_{\lambda_2}\geq0$, & 
 $y_{\lambda_2,\lambda_b} - y_{\lambda_2}\geq0$ & \\
 $y_{\lambda_3,\lambda_a} - y_{\lambda_3}\geq0$, & 
 $y_{\lambda_3,\lambda_b} - y_{\lambda_3}\geq0$ & \\
 $y_{\lambda_4,\lambda_a} - y_{\lambda_4}\geq0$, & 
 $y_{\lambda_4,\lambda_b} - y_{\lambda_4}\geq0$ & \\
 $y_{\lambda_5,\lambda_a} - y_{\lambda_5}\geq0$, & 
 $y_{\lambda_5,\lambda_b} - y_{\lambda_5}\geq0$ & \\
& & \\
\hline
\end{tabular}
\end{center}
\end{footnotesize}
\end{table}

The algorithm guarantees that there is a vertex $\lambda_i$ that satisfies
\begin{footnotesize}
$
	x_{\lambda_i} + \sum_{\lambda_j \in
	\mbox{Nbr}[\lambda_i]}{x_{\lambda_j} \leq 4},
$
\end{footnotesize}
and all vertices are searched to find such a vertex.  We then update
$\sigma(\lambda_j)$ for
vertices that are neighbors of $\lambda_i$:
\begin{footnotesize}
$
\sigma_{\mbox{new}}(\lambda_j) = \sigma(\lambda_j) - \sigma(\lambda_i), \quad \quad \mbox{ if $\lambda_j \in \mbox{Nbr}[\lambda_i]$}
$
\end{footnotesize}
Vertices that  have no neighbors, such as  $\lambda_1$ and $\lambda_5$
in our example, are included in our final solution.  Using the updated
values  of  $\sigma(\lambda_i)$, we  remove  vertices of  substructure
pairs that have score less than 0, and recursively solve the (smaller)
LP problem again.   The selected final set of  fragments are then used
to  construct  a  substructure  of  $\cS_b$.  They  are  then  aligned
structurally by minimizing cRMSD.

\paragraph{Similarity score.}
The similarity score $ \sigma(\lambda) = \sigma(\lambda_a, \lambda_b)$
between two aligned substructures $\lambda_a$ and $\lambda_b$ is a
weighted sum of a shape similarity value derived from cRMSD value 
and 
a sequence composition score (SCS):
$
        \mbox{SCS} = \sum_{i=1}^N{B(A_{a,i},A_{b,i})},
$
where $A_{a,i}$ and $A_{b,i}i$ are the amino acid residue type at
aligned position $i$, $N$ the number of residues in the aligned fragment, and
$B(A_{a,i},A_{b,i})$ the similarity score between $A_{a,i}$ and
$A_{b,i}$ by the modified {\sc Blosum50} matrix, 
in which a constant is added   
to all entries such that the smallest entry is 1.0.
The  similarity score
$\sigma(\lambda_a, \lambda_b)$ is calculated as follows: \\
\begin{footnotesize}
\[
\sigma(\lambda_a, \lambda_b) = 
\frac{\alpha(C-\mbox{cRMSD}) + \beta(\mbox{SCS})}{N},
\]
\end{footnotesize}
For the current implementation,
$C = 60, \alpha=6$, and $\beta = 1$.

To improve computational efficiency, a threshold measure is introduced
to immediately exclude low scoring fragment pairs from further
consideration.  Only fragment pairs scoring above the threshold will
be candidates to be included in the final solution.  In the above
example, we assume that only fragment pairs represented by vertices
$\lambda_1$, $\lambda_2$, $\lambda_3$, $\lambda_4$, and $\lambda_5$
(Figure~\ref{Fig:conflict_1}b) are above the threshold.

\subsubsection*{\center {\sc Results}}
We discuss the structural alignments of several well known examples of
naturally  occurring  circularly permuted  proteins.  

The results including other examples are summarized in
Table~\ref{Tab:benchmark}, which includes two additional examples of
experimentally constructed human-made permuted proteins.  None of them
are found by existing servers.

\begin{table}[tb]
        \begin{footnotesize} 
 	\begin{center} 
	\caption{\sf \footnotesize 
	 Results from  the structural alignment  of circularly permuted
         protein, where  $N$ equal the  number of aligned  residues in
         the final  solution. Human-made circularly  permuted proteins
         are listed at  the bottom separated by a  line from naturally
         occurring proteins.  }  \label{Tab:benchmark}
\vspace*{0.1in}
\begin{tabular}{|ll|cc|l|}
\hline
PDB/Size   & PDB/Size   & Frag- & $N$ & cRMSD\\
           &            & ments &     &      \\
\hline
1rin/180   & 2cna/237   & 3     & 45  & 0.877\\
1rsy/121   & 1qas/123   & 4     & 44  & 1.107\\
1nkl/78    & 1qdm/74    & 6     & 48  & 1.832\\
1onr/316   & 1fba/360   & 7     & 77  & 2.444\\
1aqi/191   & 1boo/259   & 4     & 66  & 3.571\\
\hline
1avd/123   & 1swg/112   & 6     & 66  & 0.815\\
1gbg/214   & 1ajk/212   & 5     & 110 & 0.347\\
\hline
\end{tabular}

         \end{center}
         \end{footnotesize}
\end{table}

\begin{figure}[bt]
       \centerline{\epsfig{figure=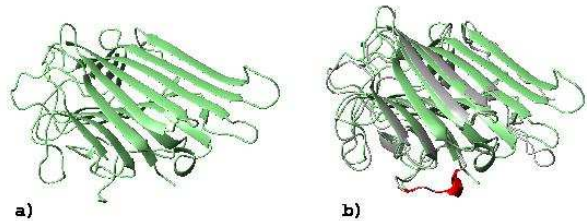,width=80mm}}
       \caption{\sf \footnotesize The  structure  of  concanavalin A  ({\tt  2cna})
       (a) and its superposition to lectin from garden
       pea  ({\tt 1rin})  (shown in  gray) (b).   The
       residues spanning the break due to the circular permutation are
       highlighted in red.} 
       \label{Fig:lectins}
\end{figure}

The first naturally occurring circular permutation was found in
concanavalin A.  and lectin favin.  The structures of lectin from
garden pea ({\tt 1rin}) and concanavalin A ({\tt 2cna})
(Figure~\ref{Fig:lectins}a,b).  In the structural alignment, three
fragments align over 45 residues with a root mean square distance of
0.82\AA.  The superposition based on the aligned fragments is shown in
Figure~\ref{Fig:lectins}b.  The residues spanning a break in the
backbone due to the permutation are highlighted in red.

\subsubsection*{{\sc Discussion}} 
The  approximation algorithm  introduced in  this work  can  find good
solutions for the problem of detecting circular permuted proteins.  In
contrast to methods based on  sequence alignment alone, the ability to
incorporate both  structural and sequence similarity  is critical.  An
experimentation  in  the  alignment  of  {\tt  1rin}  and  {\tt  2cna}
illustrates this point.  After turning off the contribution from cRMSD
in the  similarity function $\sigma(\lambda_a,  \lambda_b)$ by setting
$\alpha = 0$,  we find altogether 452 fragments that  score in the top
10\% of the  set of all SCS  scores.  This number is too  large as the
size  of the  conflict graph  will  be large.   This makes  subsequent
comparison  very  difficult.  Similarly,  cRMSD  measurement alone  is
inadequate for  comparison. 
When matching fragments from {\tt 1rin} and {\tt 2cna}, we find there
are 287 fragments that can be aligned with an cRMSD value $< 3.0$ \AA
after setting $\beta=0$.  The use of composite similarity function is
therefore essential to reduce false positives in substructure
alignments.

In our method, scoring function plays pivotal role in detecting
substructure similarity of proteins.  Much experimentations are needed
to optimize the similarity scoring system for general sequence order
independent structural alignment.  A necessary ingredient for a fully
automated search of circular permuted proteins in database is a
statistical measurement of significance of matched substructures.
Since cRMSD measurement and therefore $\sigma(\lambda_a, \lambda_b)$
strongly depends on the length of matched parts and the gaps of the
sequence break, a statistical model assessing the $p$-value of a
measured $\sigma(\lambda)$ against a null model needs to be
developed. Circular permutations that are likely to be biologically
significant can then be automatically declared \cite{pvsoar03}.

{\small
\begin{spacing}{.7}
\bibliography{align,add,liang,mathbio,pvSoar,potential}
\bibliographystyle{unsrt}
\end{spacing}
}

\end{document}